\documentclass[letterpaper,12pt]{article}
\usepackage{calc,layout,epsfig,setspace,palatino,booktabs,rotating,multirow,amsmath,hyperref}
\title{A New Concept for Searching for Time-Reversal Symmetry Violation Using Pa-229 Ions Trapped in Optical Crystals}
\author{Jaideep Singh\footnote{singhj AT nscl DOT msu DOT edu}\\ Spinlab: \href{https://spinlab.me}{http://spinlab.me} \\ National Superconducting Cyclotron Laboratory\\ Michigan State University}
\date{\today}
\normalsize
\begin{document}

\maketitle

\begin{abstract}
Certain pear-shaped nuclei are expected to have enhanced sensitivity to time-reversal and parity-violating interactions originating within the nuclear medium. In particular, Protactinium-229 is thought to be about 100,000 times more sensitive than Mercury-199 which currently sets some of the most stringent limits for these types of interactions. Several challenges would first have to be addressed in order to take advantage of this discovery potential. First, there is not currently a significant source of Pa-229 (1.5 day half-life); however, there are plans to harvest Pa-229 at the Facility for Rare Isotope Beams at Michigan State University. Second, the spin-5/2 nucleus of Pa-229 limits its coherence time while also making it sensitive to systematic effects related to local electric field gradients. On the other hand, this also give Pa-229 an additional source of signal in the form of a magnetic quadrupole moment (MQM) which violates the same symmetries as an EDM but is not observable in spin-1/2 systems. Third, in order to compensate for the small atom numbers and short coherence times, the Pa-229 atoms would have to be probed with exceptionally large electric and magnetic fields that may be possible if Pa-229 ions are embedded inside an optical crystal. We will describe some aspects of this concept using the stable Praseodymium-141 isotope as a surrogate which has the same nuclear spin and similar atomic structure of Pa-229.
\end{abstract}

\section{Electric Dipole Moments and CP-Violation}
Why is there more matter than antimatter in the observable Universe?
Based on arguments laid out by Sakharov \cite{Sakharov}, the answer to this question requires, among other things, the existence of charge-conjugation-parity ($CP$) violating interactions.
Although $CP$-violation has been observed in the rare decays of kaons \cite {cp-kaon} and B mesons \cite{babar,belle} mediated by the weak interaction, the amount observed is far too feeble to explain the observed baryon asymmetry \cite{huet-sather}.
Many extensions to the Standard Model (SM), such as Supersymmetry, naturally allow for new sources of $CP$-violation \cite{ritzpop}.
The existence of a permanent electric dipole moment (EDM) of a particle indicates both time-reversal symmetry violation ($T$) and parity violation ($P$).
By the $CPT$-theorem, $T$-violation implies $CP$-violation and, consequently, a non-zero EDM is an indication of $CP$-violation as well.
The $CP$-violation currently encoded in the Standard Model by the Cabibbo-Kobayashi-Maskawa (CKM) matrix \cite{pdg} generates EDMs that are too small to be observed for the foreseeable future \cite{ermk13}.
Because they are effectively free from Standard Model ``backgrounds,'' EDMs are a particularly clean signature of $CP$-violation from physics Beyond the Standard Model (BSM).
EDM searches are complementary to $CP$-violation searches at the Large Hadron Collider (LHC) \cite{Brod} and generically probe energy scales that are beyond the reach of LHC.
Both the need for and the predictions of additional sources of $CP$-violation generate the significant discovery potential that motivates the world-wide search for EDMs of protons, light nuclei, neutrons, muons, paramagnetic atoms \& molecules, and diamagnetic atoms \& molecules \cite{rmp2019}.

\section{Complementary Searches For Different Sources of CP-Violation}
EDM searches in different systems are complementary since they are sensitive to a different linear combination of a variety $CP$-sources \cite{global,fj18}.
Neutron EDM \cite{nedm} is mostly sensitive to a ``short range'' intrinsic neutron EDM as well as isoscalar \& isovector $CP$-violating nucleon-nucleon interactions.
Paramagnetic atoms (Tl \cite{regan}) and molecules (YbF \cite{ybf}, ThO \cite{tho}, HfF$^+$ \cite{hffp}) have the appearance of a single unpaired electron and are mostly sensitive to an intrinsic electron
EDM and nuclear spin-independent $CP$-violating electron-nucleus interactions.
In paramagnetic systems, the contribution of an electron EDM is amplified by the relativistic atomic structure of heavy alkali-like atoms \cite{sandars,ignatovich}.
Diamagnetic atoms ($^{199}\mathrm{Hg}$ \cite{graner}, $^{129}\mathrm{Xe}$ \cite{xe129}, $^{225}\mathrm{Ra}$ \cite{bishof}) and molecules (TlF \cite{tlf}) have the appearance of a partially screened nucleus and are mostly sensitive to isoscalar \& isovector $CP$-violating nucleon-nucleon interactions and nuclear spin-dependent $CP$-violating electron-nucleus interactions.
In diamagnetic systems, the nuclear EDM is almost completely screened by the surrounding electron cloud \cite{schiff}.
This screening is not perfect in heavy atoms because of their relativistic atomic structure as well as the finite size of the nucleus.
The residual effect from the nucleus that contributes to the atomic EDM is called the Schiff moment. 

\section{Diamagnetic Systems and Octupole Enhancements}
\begin{figure}
        \centering        
       	 \epsfig{file=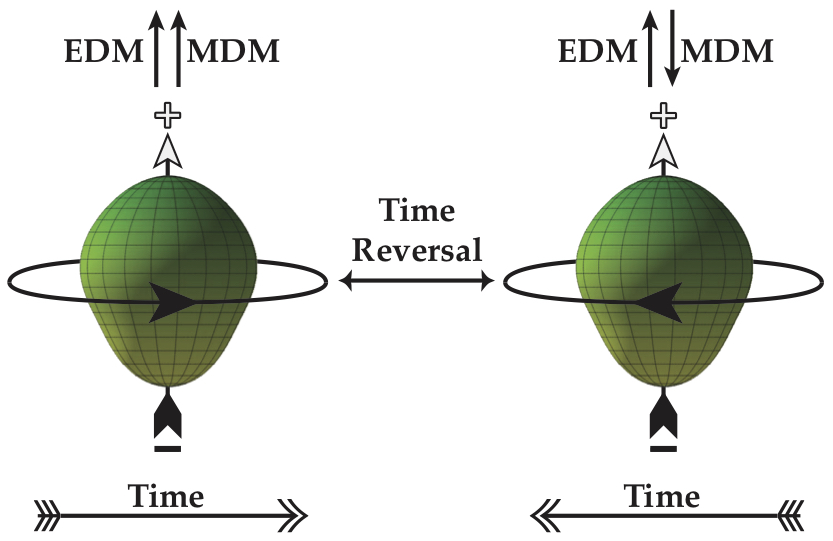,height=2.3in}
	 \caption{Octupole-Deformed Nucleus with Spin in its Body Frame Under Time Reversal\label{fig:pear}. EDM (MDM) is the electric (magnetic) dipole moment. An octupole deformation has a reflection-asymmetric ``pear'' shape.}
\end{figure}

For diamagnetic systems, Mercury-199 \cite{graner} currently sets the most stringent limits on $CP$-violating interactions originating within the nuclear medium.
An attractive alternative are diamagnetic atoms that have pear-shaped (octupole-deformed) nuclei such as Radium ($^{225}\mathrm{Ra}$), Radon ($^{221,223}\mathrm{Rn}$), and Protactinium ($^{229}\mathrm{Pa}$) \cite{gaffney13,ahmad82}.
The observable effect of the same underlying sources of $CP$-violation is amplified by orders of magnitude in comparison to $^{199}\mathrm{Hg}$.
A nucleus with an octupole deformation, see Fig.~(\ref{fig:pear}), is expected to have an energy level structure which includes a ladder of nearly degenerate parity doublets \cite{feinberg77,hh83}.
Each nuclear state of the parity doublet has a well-defined parity and is a linear combination of nuclear states that have large intrinsic body-frame octupole deformations and Schiff moments.
In the presence of a parity-violating nucleon-nucleon interaction, the two states that make up the parity doublet mix and results in an enhanced lab-frame Schiff moment \cite{spevak96}.
The contribution to the atomic EDM due to this Schiff moment scales in the following way \cite{spevak97,dzuba2002}:
\begin{equation}
d_\mathrm{atom} \propto \frac{\beta_2 \left < \beta_3^2 \right >  Z^3 A^{2/3}}{\Delta E}
\end{equation} 
where $\beta_2$ is the quadrupole deformation parameter, $\beta_3$ is the octupole deformation parameter, $Z$ is atomic number, $A$ is the atomic weight, and $\Delta E$ is the energy difference between the parity doublet. 
In very round numbers when compared to $^{199}\mathrm{Hg}$ \cite{ban10}, the enhancement for $^{221/223}\mathrm{Rn}$ is about $10^2$, for $^{225}\mathrm{Ra}$ is about $10^3$ \cite{engel05}, and for $^{229}\mathrm{Pa}$ is about $10^5$ \cite{vvf2008}.  
The particularly large enhancement that is expected for $^{229}\mathrm{Pa}$ is due to the unusually small energy difference which may be as small as $\Delta E = 0.06\ \mathrm{keV}$ \cite{ahmad15}. 
More details are listed in Tab.~(\ref{tab:isotope}).
\begin{table}
\centering
\begin{tabular}{ccccccccc}
\toprule \addlinespace
species		&	\% N.A. &	spin	&	$Z$	&	$A$		&	$\Delta E $	&	sensitivity	&	$2\sigma$ limit  	&	ref.	 \\
		&	$\tau_{1/2}$(FRIB)	&		&		&		& (keV)	&	rel. to Hg	&	$(e\ \mathrm{cm})$ 	&	
\\ \addlinespace \midrule \addlinespace
Xe	    &	26.4		&	1/2				&	54	&	129		&	-									&	0.1		&	$6.6\!\times\!10^{-27}$ &  \cite{xe129}	\\
Pr	    &	100		&	5/2				&	59	&	141		&	-									&	0.2		&	-					    &  -	\\
Yb		&	14.3		&	1/2				&	70	&	171		&	-									&	0.6	    &	-	                    &  -	\\
Yb		&	16.1		&	5/2				&	70	&	173		&	-									&	0.6	    &	-	                    &  -	\\
Hg		&	16.9		&	1/2				&	80	&	199		&	-									&	1.0	    & $7.4\!\times\!10^{-30}$   &  \cite{graner}	\\ \addlinespace
Rn		&	26 m ($10^9$/s)		&	7/2				&	86	&	221	    &	100?						 	    			& $10^2$	& -                         &  - \\
Rn		&	24 m ($10^9$/s)		&	7/2				&	86	&	223	    &	100?					 	    			& $10^2$	& -                         &  - \\
Ra      &   11.4 d ($10^9$/s)     &   3/2             &   88  &   223     &   50 											& $10^3$    & -                         &  - \\ 
Ra	    &	14.9 d	($10^9$/s)	&	1/2				&	88	&	225		&	55										& $10^3$	&	$1.4\!\times\!10^{-23}$	&	\cite{bishof}	\\
Pa	    &	1.50 d	($10^{10}$/s)	&	5/2				&	91	&	229	    &	0.06?										& $10^5$	&	-	                    &  -	\\
\addlinespace \bottomrule
\end{tabular}
\caption{Data Table for Candidate EDM Isotopes. \label{tab:isotope} FRIB refers to the projected production rate at the Facility for Rare Isotopes Beams at Michigan State University (East Lansing, MI, USA). The sensitivity relative to Hg-199 are estimates using \cite{ban10,dekb18} that are probably reliable to within a factor of three for the stable isotopes and within an order of magnitude for the octupole deformed isotopes. Most of the difference in sensitivity for octupole species is due to differing values for $\Delta E$.}
\end{table}

\section{Fundamental Statistical Sensitivity of EDM Experiments}
A common measurement principle behind several EDM searches is to perform a low-field nuclear spin precession experiment and to search for a small electric-field correlated shift in the Larmor frequency.
In the presence of parallel electric and magnetic fields, the Larmor frequency or spin-precession frequency of a spin-$1/2$ particle is given by:
\begin{equation}
	\nu_\pm = \frac{2 \mu B_\pm}{h} \pm \frac{2 d E_\pm}{h}
\end{equation}
where $d$ ($\mu$) is the particle's electric (magnetic) dipole moment EDM (MDM), $E_\pm$ ($B_\pm$) is the electric (magnetic) field, $h$ is the Planck constant, and $+(-)$ is for when the $E$- \& $B$-fields are parallel (anti-parallel).
In order to isolate the EDM, we take the difference in the two frequencies to get:
\begin{equation}
\Delta \nu = \frac{4 d \bar{E}}{h} + \frac{2 \mu \Delta B}{h} 
\end{equation}
where $\bar{E}$ is the magnitude of the electric field averaged over the two frequency measurements and $\Delta B$ is the difference in the magnetic field between the two frequency measurements.
If the magnetic field is perfectly stable and uniform, then the frequency difference is directly proportional to the EDM and the statistical uncertainty 
of the measurement is given solely by the uncertainty in the frequency measurement, which ideally is just the inverse of the spin precession observation time $\tau$ \cite{cn68}:
\begin{equation}
	\sigma_\nu = \frac{1}{2 \pi \tau} \rightarrow \frac{\sigma_d}{\sqrt{N_m}} = \frac{\hbar}{2 \bar{E} \sqrt{\epsilon N_a T \tau} }  
\end{equation}
where $N_m$ is the number of frequency difference measurements, $N_a$ is the total number of particles probed, $T$ is the total integration time, and $\epsilon$ is the experimental efficiency.
It is apparent that maximizing the statistical sensitivity involves performing the experiment at large $E$-fields, at high experimental efficiency, with large particle numbers, long
integration times, and long spin precession observation times.
The experimental efficiency depends critically on the experimental geometry, specifics of the EDM-correlated observable, and, in particular, the state preparation and readout schemes.
Any variations in the $B$-field reduces the statistical sensitivity of the experiment. 
Furthermore, any change in the $B$-field that is correlated to the $E$-field polarity would create a large systematic frequency shift that would mask the EDM signal.
An experimental approach for which the $E$-field polarity need not be toggled would be particularly immune to a class of systematic effects.

\section{Opportunities For a Next Generation EDM Search in the FRIB Era}
When dealing with rare isotopes, the major challenge is achieving near unity experimental efficiency in terms of both using up all of the atoms before they decay away and the detection scheme.
In order to eventually carry out a test of time-reversal symmetry with unprecedented sensitivity, 
new experimental techniques, from isotope production to observable measurement, need to be explored and developed first. 
In particular, several challenges would have to be overcome in order to realize a competitive $^{229}\mathrm{Pa}$ experiment. 
First, because of its 1.5 day half-life, there is not currently a significant source of $^{229}\mathrm{Pa}$.
Within the next decade, the Facility for Rare Isotope Beams (FRIB) at Michigan State University (East Lansing, MI, USA) will provide a steady supply of large quantities of several promising EDM candidate isotopes and 
some, such as Pa-229, will be available in sufficiently large quantities for the first time.
In general, a more regular and frequent supply of these isotopes will be critical for studying the systematics of EDM searches.

Second, the spin-5/2 nucleus of $^{229}\mathrm{Pa}$ potentially limits its coherence time while also making it sensitive to systematic effects related to local field gradients. 
On the other hand, this also gives $^{229}\mathrm{Pa}$ an additional source of signal in the form of a magnetic quadrupole moment (MQM) which violates the same symmetries as an EDM but is not observable in spin-1/2 systems. 
Third, in order to compensate for the small atom numbers and potentially short coherence times, the $^{229}\mathrm{Pa}$ atoms would have to be probed with exceptionally large electric \& magnetic fields that are only possible if $^{229}\mathrm{Pa}$ is a part of a polar molecule such as PaO or PaN or embedded inside of optical crystal such as $\mathrm{YSiO_4}$. 
Finally, systematic effects would have to be controlled at the same level or better than the desired statistical sensitivity. 

We are proposing to manipulate nuclei embedded inside an optically transparent solid at cryogenic temperatures.
Implantation into a solid is potentially an effective way to efficiently capture and repeatedly probe all of the small number of nuclei. 
An optically transparent host medium at cryogenic temperatures would allow for the laser manipulation of the nuclei in a thermally quiet and stable environment.
Since the optical manipulation of nuclear spins must be facilitated via a hyperfine interaction, the nuclei will be located within a molecule embedded in a rare gas solid \cite{vhh18} or an atomic ion inside of an optical crystal. 
In such systems, the nuclei are exposed to extraordinarily large electric fields and magnetic field gradients, which significantly amplify the measurability of the time-reversal violating effects thereby compensating for the small number of nuclei.
Finally, the spectra of such systems are often unusually rich and complex, which could allow for the possibility of internal reversals that would allow for exquisite control of systematic effects. 
A promising approach is to use $^{229}\mathrm{Pa}$ ions embedded inside of optical crystals, which are expected to have the same favorable optical properties as the lanthanide ions.

\section{Rare Earth Ions Embedded in Cryogenic Optical Crystals}
Rare earth lanthanide ions in optical crystals have laser-friendly transitions and narrow optical linewidths at cryogenic temperatures \cite{rei}. 
While the inhomogeneous linewidths are 1 to 10 GHz at cryogenic temperatures, the homogenous linewidths vary from 10 Hz to 0.1 MHz.
For this reason, these systems have been studied extensively over the last fifty years as both a laser medium and for quantum information processing.
They offer five potential advantages for EDM searches.

First, the linewidths are narrow enough for the hyperfine structure to be resolved.
This means that the nuclear spins can be optically manipulated coherently \cite{longdell05}.
Optically detected NMR \& NQR has been widely used in rare earth spectroscopy and some examples include measurements of the hyperfine structure of $\mathrm{Pr^{3+}}$ in $\mathrm{LaF_3}$ \cite{eri77} and $\mathrm{Y_2SiO_5}$ \cite{ecm95}, a high precision measurement of the nuclear magnetic moment of $\mathrm{^{141}Pr}$ \cite{mbs82}, and measurements of the nuclear spin relaxation times in the electronic ground state of $\mathrm{Pr^{3+}}$ in $\mathrm{YAlO_3}$ \cite{kli03}. 
Second, depending on the symmetry of the trapping site of the ion, there is a large local electric field (1-10 MV/cm)\cite{macf2007}.
Third, they can be fabricated via ion implantation with efficiencies of 50\% or higher \cite{kornher16}.
Fourth, single ion detection has been demonstrated \cite{kol12}, which may be needed to make most efficient use a small number of atoms.
Finally, because they are considered promising candidates for nuclear spin-based optically-addressable memory for quantum 
information processing, techniques for extending coherence times to as long as hours have already been demonstrated \cite{zhong}. 
Putting this altogether provides the promising path to search for the EDM of Protactinium-229, which is a heavy actinide version of the lanthanide ions which exhibit these desirable properties.

\section{EDM Searches in Optical Crystals}
The earliest searches for time-reversal violation using ions in a solid involved measuring the degree of non-degeneracy of a Kramer's doublet \cite{SACHS1,SACHS2,bro61,rb63}, where, in a perfectly time-reversal invariant system, the energy levels of a Kramer's doublet are exactly degenerate. 
In the case of the Royce \& Bloembergen experiment \cite{rb63}, the EDM observable was a shift in the linear ``Stark'' splitting  of the EPR frequency of $\mathrm{Cr^{3+}}$ ions in $\mathrm{Al_2O_3}$ upon reversal of a magnetic field. 
The key aspect of this experiment is that the $\mathrm{Cr^{3+}}$ ions sit in one of two distinct sites that are related to each other by a spatial inversion. 
The local crystal field at these two sites are equal in magnitude but opposite in direction.
The importance of such sites is that they provide a built-in reversal without the need to alter the external $E$-field which 
provides a powerful way to reject a certain class of $E$-field correlated systematic effects.  

It is worth expanding on this last point in more detail.
An externally applied $E$-field distorts the crystal field around an ion.
As a consequence, the ion spin parameters in the Hamiltonian, such as the g-factor, nuclear quadrupole coupling constant, and hyperfine coupling constant are modified \cite{ksw61,kiel66,lucken,mims}. 
The modification is due to an interference term between the externally applied field and an internal crystal field.
In crystals that lack inversion symmetry, this internal crystal field is odd.
This implies that the spin parameters are shifted opposite in sign for ions occupying the two different sites.
Therefore, the application of an external $E$-field can be used to spectroscopically distinguish the two crystal sites which have equal but opposite local crystal fields.

Since the dominant forces acting on the ion in a solid are electrostatic in nature, the net electric field at the ion is, to zeroth order, nothing.
However, if the electrostatic forces on the ion due to the crystal are partially balanced by magnetic forces due to spin-orbit coupling, then there is potentially a net nonzero crystal field at the location of the ion even in the absence of an external $E$-field.
The size of this residual crystal field can be estimated by $(\lambda/\Delta)(1\ \mathrm{V}/1\ \mathrm{\AA})$, where $\lambda$ is the spin-orbit splitting and $\Delta$ is the energy difference between the ground state electronic state and the closest excited state with the same spin but opposite parity \cite{rb63}. 
For rare earth actinides (lanthanides), the ratio $\lambda/\Delta$ is of order $0.1$ (0.01) resulting in residual crystal fields of order $10^7\ \mathrm{V/cm}$ ($10^6\ \mathrm{V/cm}$).  

\begin{figure}
        \centering        
       	 \epsfig{file=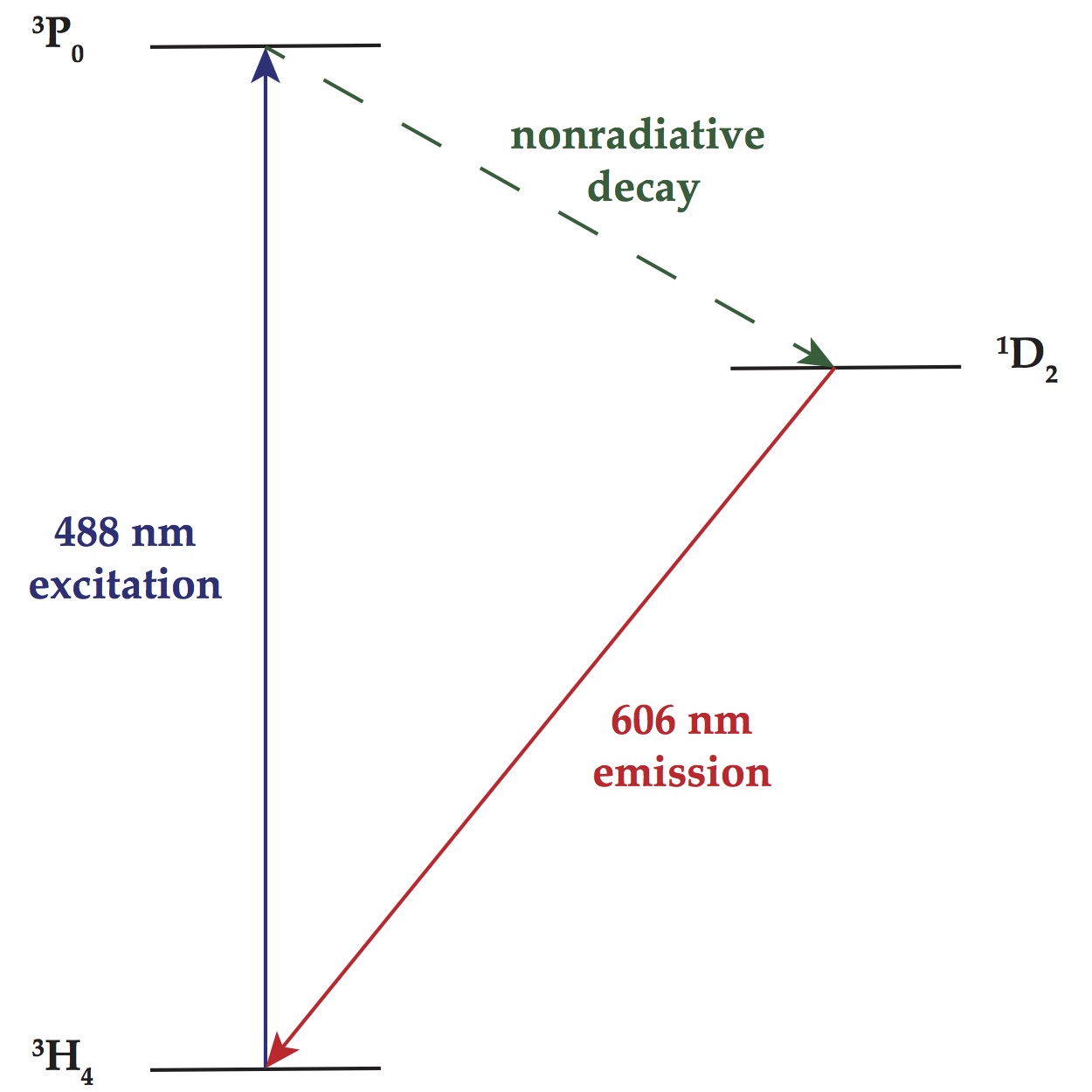,height=2.3in}
        \epsfig{file=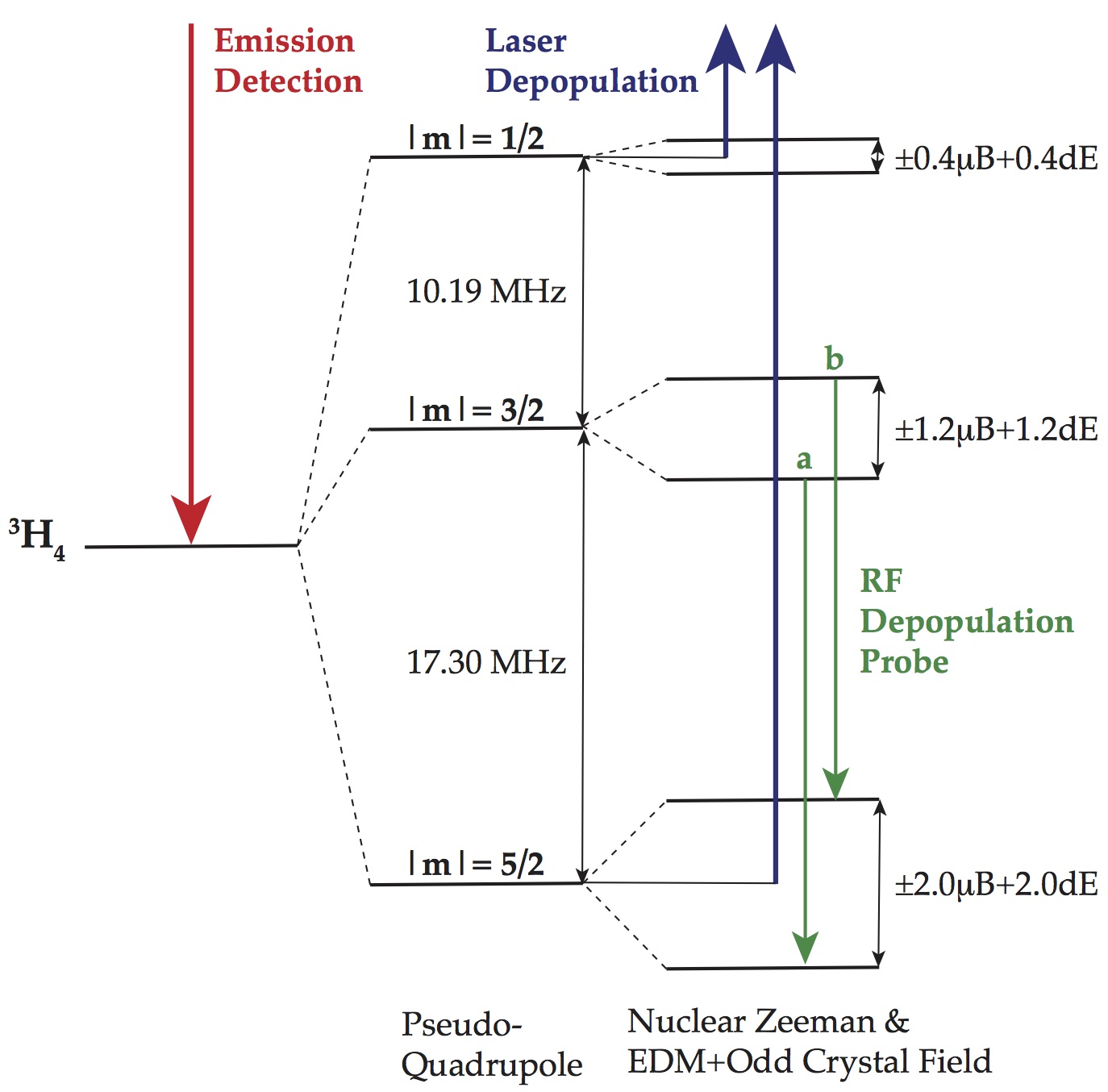,height=2.3in}
        \caption{\bf Left.\rm\ Simplified Excitation Cycle of $\mathrm{^{141}Pr^{3+}}$ in $\mathrm{Y_2SiO_5}$ from the $\mathrm{^3H_4}$ ground state to the $\mathrm{^3P_0}$ excited state. 
        The down conversion of the 488 nm excitation light to 606 nm emission light allows us to optically probe the $\mathrm{^3H_4}$ ground state populations. \bf Right.\rm\ $\mathrm{^3H_4}$ Ground State Dynamics of $\mathrm{^{141}Pr^{3+}}$ in $\mathrm{Y_2SiO_5}$. Site selective spectral hole-burning is used to depopulate the $\pm1/2$ \& $\pm5/2$ doublets and populate the $\pm3/2$ doublets. Each doublet is split by the presence of an externally applied magnetic field ($B$) and possibly a permanent EDM ($d$) coupled to an internal crystal field ($E$). The size of the splitting can determined by applying an RF field at a frequency that transfers atoms from the $\pm3/2$ doublet to the $\pm5/2$ doublet. This will results in excitation from the $\pm5/2$ doublet to the $\mathrm{^3P_0}$ excited state which can be monitored by the presence of additional 606 nm emission light. The difference in RF probe frequencies between the transitions labeled ``a'' and ``b'' isolates the combined splitting due to the external magnetic and internal electric fields. The two sites, with opposite pointing internal $E$-fields, can be spectroscopically distinguished upon application of an external $E$-field. The EDM contribution is then isolated by finding the shift in this splitting when the external magnetic field is reversed with respect to the internal electric field for the two different sites. The accuracy \& precision of the determination of the EDM contribution depends crucially on the exactness \& repeatability of the magnetic field reversal.  \label{fig:pr141} }
\end{figure}

\section{Outlook \& Conclusion}
Several precursor measurements are needed before launching a search for time-reversal violation using Pa-229 ions in optical crystals.
First, a sensitive spin precession frequency readout scheme should be demonstrated in a stable surrogate such as Pr-141, which has the same nuclear spin and similar atomic structure as Pa-229, see for example Fig.~(\ref{fig:pr141}).
Once demonstrated, Pr-141 ions embedded within the same crystal as Pa-229 could be used as a magnetometer to control \& monitor the $B$-field reversal.
Second, detailed optical spectroscopy of longer-lived Pa isotopes will have to be performed in a variety of optical crystals, which will be modified from Pr spectra due to the stronger spin-orbit coupling \cite{ML09}.
Third, the relevant crystal field parameters necessary for calculating the effective internal electric field will have to be deduced from both theory and experiment.
Finally, a determination of the nuclear structure parameters needed to calculate the intrinsic enhancement factors, such as those discussed in \cite{ahmad15}, will have to be made.

Towards these ends, an ion source based on electrospray ionization \cite{kt93} for producing rare earth ions for implantation is currently under development.
These sources \cite{shaffer98,Ibrahim2006,page2008,Cox2014} offer the potential for near unity extraction of ions from a solution \cite{mar10,Cox2015} making highly efficient use of the small rare isotope atom numbers expected to be harvested from FRIB.
Assuming an internal electric field of $10^7$ V/cm, an experimental efficiency of 10\%, an integration time of 1 day, a spin coherence time of $100\ \mu$s, and an atom number of $10^{7}$, the Pa-229 EDM statistical sensitivity would be $10^{-26}\ e\ \mathrm{cm}$.
With an octupole-enhanced physics sensitivity of $10^5$ more than Hg-199, this would be the equivalent of a Hg-199 EDM measurement at the level of $10^{-31}\ e\ \mathrm{cm}$ which would be factor of 10 better than the present limit.
A key improvement to the sensitivity of this approach would be longer spin coherence times, which at high $B$-field, have been shown to be hours long \cite{zhong}.
Realizing such an improvement would provide unprecedented sensitivity to $CP$-violation in the hadronic sector.
This material is based upon work supported by the U.S. Department of Energy, Office of Science, Office of Nuclear Physics, under Award Number DE-SC0019015.


%
%

\end{document}